\documentclass[twocolumn,prb,superscriptaddress,aps,floatfix]{revtex4}

\usepackage{lscape,floatflt,graphicx}

\begin{document}
\title{Effects of morphology on phonons of nanoscopic silver grains}

\author{Gustavo A. Narvaez} \altaffiliation[Current address: ]{National
Renewable Energy Laboratory, Golden, Colorado 80401.}  \affiliation{Department
of Physics, The Ohio State University, Columbus, Ohio 43210}

\author{Jeongnim Kim} \affiliation{Materials Computation Center, University of
Illinois, Urbana-Champaign, Illinois 61801} \affiliation{National Center for
Supercomputer Applications, University of Illinois, Urbana-Champaign, Illinois
61801}

\author{John W. Wilkins} \affiliation{Department of Physics, The Ohio State
University, Columbus, Ohio 43210}

\date{\today}

\begin{abstract}
  The morphology of nanoscopic Ag grains significantly affects the phonons.
  Atomistic simulations show that realistic nanograin models display complex
  vibrational properties. (1) Single-crystalline grains.  Nearly-pure
  torsional and radial phonons appear at low frequencies. For low-energy,
  faceted models, the breathing mode and acoustic gap (lowest frequency) are
  about 10\% lower than predicted by elasticity theory (ET) for a continuum
  sphere of the same volume.  The sharp edges and the atomic lattice split the
  ET-acoustic-gap quintet into a doublet and triplet.  The surface protrusions
  associated with nearly spherical, high-energy models produce a smaller
  acoustic gap and a higher vibrational density of states (DOS) at frequencies
  $\nu<2\,{\rm THz}$. (2) Twined icosahedra. In contrast to the single-crystal
  case, the inherent strain produce a larger acoustic gap, while the core
  atoms yield a DOS tail extending beyond the highest frequency of
  single-crystalline grains. (3) Mark's decahedra, in contrast to (1) and (2),
  do not have a breathing mode; although twined and strained, do not exhibit a
  high-frequency tail in the DOS. (4) Irregular nanograins. Grain
  boundaries and surface disorder yield non-degenerate phonon frequencies, and
  significantly smaller acoustic gap.  Only these nanograins exhibit a
  low-frequency $\nu^2$ DOS in the interval 1-2${\rm THz}$.
\end{abstract}

\maketitle

%
%
\section{Introduction}

Recent experiments have probed the size dependence of the electronic-energy
relaxation rate in nanoscopic metal grains.
\cite{arbouet_PRL_2003,hodak_JCP_2000,smith_CPL_1997} Perhaps the most
striking aspect of these experiments is the finding of Arbouet and coworkers
which shows that the electron-phonon relaxation rate $\tau_{ep}$ does not
depend on the environment of the grains.\cite{arbouet_PRL_2003} Gold and
silver grains suspended in solution present the same size dependence of
$\tau_{ep}$ than those embedded in a solid matrix (e.g., Al$_2$O$_3$).

To explain the data, current models for the energy relaxation of hot electrons
due to electron-phonon coupling in nanoscopic grains rely on phenomenological
parameters and simplifications. (For example, replacing the vibrational
density of states with the Debye model.)  One has to keep in mind, however,
that the experiments involved several materials (Ag, Au, Pt) with a wide range
of grain sizes, where a myriad of stable low-energy morphologies has been
observed and proposed.
\cite{grain_geometries,baletto_JCP_2002,baletto_papers,balletto_RMP_2005}
Thus, to understand the electron-phonon coupling in nanoscopic metal grains it
would be desirable to have a description of their vibrational properties that
includes the grain's morphology: facets, grain boundaries, and surface
disorder and defects.

There are papers in the literature that have addressed the vibrational density
of states of nanoscopic metal grains using ideal, spherical models
\cite{kara_PRL_1998,sun_PRB_2001,meyer_PRB_2003} Unfortunately, the use of
these idealized models limits a detailed understanding of the vibrational
properties and the role of morphology.  Furthermore, the idealized models
neglect the influence of grain boundaries, and surface disorder and defects on
the vibrational properties of the experimental nanograins. \cite{note-0000} In
this work we focus on {\em realistic}, nanoscopic Ag grains containing
150-1400 atoms and atomistically simulate their vibrational properties, by
calculating the phonon frequencies and displacement vectors.  We show that
morphology introduces a high degree of complexity in the phonon spectra; total
and partial vibrational density of states; and phonon localization.

Our predictions, which should be general for nanoscopic metals, show the
following prominent features. (a) Low-energy, single-crystalline grains
present nearly-pure torsional and radial phonon modes.  The frequencies of the
breathing mode and the acoustic gap (lowest frequency) are nearly 10\% lower
than predicted by elasticity theory (ET) for a continuum sphere.
\cite{landau_book, patton_PRB_2003,tamura_JPC_1982} The sharp edges and atomic
lattice of the grains lead to the splitting of the acoustic gap quintet
predicted by ET in a doublet and triplet, where the magnitude of the splitting
depends on the relative number of atoms at the boundary of different facets.
High-energy, ideal spherical models present regular protrusions on the
surface.  When compared to a faceted grain of the same size, these protrusions
lead to a smaller acoustic gap, which is not necessarily degenerate; and to a
higher total vibrational density of states (DOS) at frequencies $\nu<2\,{\rm
  THz}$.

(b) As in the case of single-crystalline nanograins, twined icosahedra have a
breathing mode. Strain in these grains leads to a higher acoustic gap, and a
high-frequency tail in the DOS that extends beyond the highest frequency in
single-crystalline grains. On the other hand, twined and strained Mark's
decahedra {\em do not} have a breathing mode, neither exhibit a high-frequency
tail on DOS.

(c) Irregular nanograins with grain boundaries and surface disorder do not
have degenerate phonon frequencies, and the acoustic gap is significantly
reduced. These nanograins are the only ones to exhibit $\nu^2$ DOS in the
interval 1-2${\rm THz}$. The extent of this region depends on the nature of
disorder.

To summarize, our simulations show the complexity of the vibrational
properties of nanoscopic metals, and point out similarities between the
vibrational density of states of nanograins with grain boundaries and surface
disorder and that of massive nanocrystalline samples. This work is a required
step towards understanding the size dependence of the electron-phonon coupling
in nanoscopic metal grains.

%
%
\section{Nanograins' MORPHOLOGY AND ENERGETICS} 
\label{Section_2}

\begin{figure}[tb]
\includegraphics[width=8.5cm]{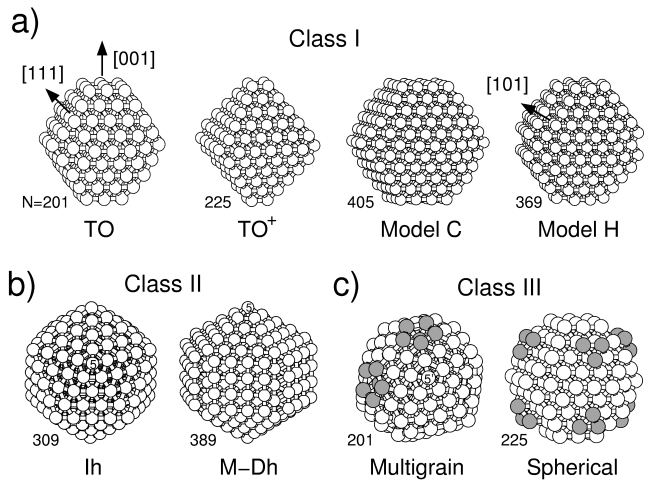}
\caption{\label{Fig_1} Classes of Ag nanograins illustrated with selected
  specimens. ``5'' [(b)] and ``$5\,^{\prime}$'' [(c)] indicate, respectively, 
a 5-fold and 5-fold-like axis. Number of atoms ($N$) is indicated. 
(a) Single-crystal grains with [111] and [001] facets in the regular ({\sf
TO}) and irregular ({\sf TO}$^{+}$) truncated octahedron geometry. Compared 
to ${\sf TO^+}$, {\sf Model C} has bigger [001] facets and similar [111]-facet
geometry. {\sf Model H} grains present [110] facets. (b) Twined icosahedron
({\sf Ih}) and Mark's decahedron ({\sf M-Dh}). (c) Multigrain nanograins
exhibit grain boundaries and surface disorder; gray atoms highlight vertex 
vacancies. Spherical nanograins result from decorating regular, faceted
specimens. Gray atoms decorating [111] facets of {\sf TO} [see (a)] yield the
shown ``sphere.''}
\end{figure}

The morphology of a nanograin is specified by the number of atoms $N$, shape
and atomic arrangement. The shape refers to the symmetry and facets, while the
atomic arrangement pertains to the nanograin's crystalline characteristics:
single-crystalline, twined, surface-disordered and defective. Here, we
classify the model nanograins---with sizes that range from N=150-1400---in
three broad classes:

(I) Single-crystalline nanograins derived from the face-centered-cubic lattice
of Ag with facets oriented in the [111], [001], and [101] directions
(depending on size). Figure \ref{Fig_1}(a) illustrates class I grains.
{\sf TO} and ${\sf TO^+}$ originate from different degrees of truncation of an
octahedron ({\sf Oh}).  {\sf TO}'s are regular in the sense that the number of
atoms in every edge is the same and the [111] and [001] facets form regular
hexagons and squares, respectively.  ${\sf TO^+}$ represents a group of grains
that result from removing the vertices of the {\sf Oh} first and then [001]
facets until reaching the {\sf TO}.  The latter protocol leads to grains that
have more atoms in the edges connecting [111] facets than in the [111]/[001]
boundary.  {\sf Model C} results from truncating cuboctahedra.  In these
grains, [001] facets are larger than the corresponding facets in ${\sf TO^+}$,
while the geometry of [111] facets is preserved. On the other hand, by simply
removing the atoms lying in the [111]-facets boundary, {\sf Model C} leads to
{\sf Model H}.  By construction, {\sf Model H} nanograins show [101] facets.

(II) Twined icosahedra ({\sf Ih}) and its variants,\cite{note-020} and Mark's
decahedra ({\sf M-Dh}).  Examples of {\sf Ih} and {\sf M-Dh} appear in Fig.
\ref{Fig_1}. Both these grains are described in detail in the literature.
\cite{yang_JCG_1979,marks_RepProgPhys_1994} {\sf Ih}'s are strained and
feature [111] facets, 12 vertices, 20 internal grains, and six 5-fold axis;
one of them appears indicated with a ``5'' in Fig. \ref{Fig_1}.  In contrast,
{\sf M-Dh}'s are less strained, have multiple faceting and a single 5-fold
axis.
 
(III) Complex, partially disordered multigrain nanograins that have grain
boundaries, surface disorder, and defects; see Fig. \ref{Fig_1}(c).  Simulated
annealing is used to generate these grains.  The most common surface defects
are vertex vacancies [shown in grey in Fig. \ref{Fig_1}(c)], terraces, and
stacking faults. In most cases, the grain boundaries lead to 5-fold-like axis
and a high number (about 10) of internal grains.  However, we have also
encountered multigrains with a single twin boundary or internal stacking
fault.

%
%

We also include in class III the so-called {\em spherical} nanograins [Figure
\ref{Fig_1}(c)]. Despite their unrealistic morphology, these grains have been
the model of choice in recent simulations of the vibrational properties of
nanoscopic metal grains.\cite{kara_PRL_1998,meyer_PRB_2003} Although
single-crystalline, these grains have small protrusions (grey atoms in the
figure) on the [111] and, depending on size, [001] facets of otherwise compact
specimens.  These protrusions act as surface defects, despite of being highly
regular. (Indeed, one may consider the compact nanograins to be {\em
  decorated} by the protrusions. These surface defects contribute to the
higher energies of the spherical nanograins relative to other structures and
are removed by simulated annealing starting at high temperatures.

%
\begin{figure}[htb]
\includegraphics[width=8.5cm]{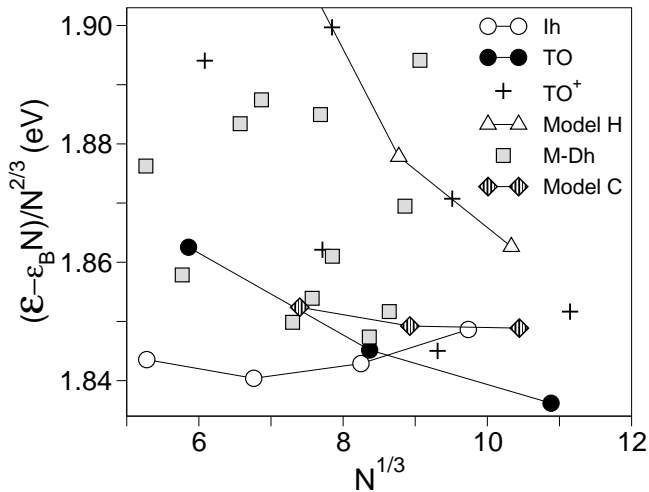}
\caption{\label{Fig_2} Morphology effects on the total energy ${\cal E}$ of
  grains in Fig. \ref{Fig_1}(a) and (b). ${\cal E}$ is measured relative to
  bulk ($\varepsilon_B=2.85\,{\rm eV}$).  Twined icosahedra ({\sf Ih}) are the
  most stable structures from 150 to $N\simeq 600$; where the cohesive
  energies of {\sf Ih}, {\sf M-Dh}, and {\sf TO} are nearly degenerate. The
  energy of ${\sf TO}^{+}$ nanograins separates in two {\em bands} (see Sec.
  \ref{Section_2}). The upper band follows the progression of {\sf Model H}.
  Increased faceting is responsible for the noticeable scattering in {\sf
    M-Dh}'s energy.}
\end{figure}
%

Figure \ref{Fig_2} shows the effect of morphology on the total energy ${\cal
  E}$ for class I and II.  {\sf Ih} are the most stable grains up to $N\simeq
600$, where the cohesive energies of {\sf Ih}, {\sf M-Dh} and {\sf TO} are
effectively degenerate. For larger grains, strain dominates over efficient
faceting and single-crystalline grains become more stable. ${\sf TO^+}$ splits
into two {\em bands}, indexed by the number of atoms in [111]-facet
boundaries; the lower and upper bands corresponding to 5 and 7 atoms,
respectively. The lower and upper band merge at $N=1289$ and $N=1975$,
respectively, with {\sf TO} energies.  {\sf Model H} tracks the ${\sf TO^+}$
upper band.
Mark's decahedra show significant fluctuations in the size dependence of
${\cal E}$ due to increased faceting; which permits generating {\sf M-Dh}
grains with significantly different {\em sphericity} even if the number of
atoms is slightly changed.\cite{note-021}

Considering the energetics of class III, Table \ref{Table_1} lists the total
energies of multigrain and spherical nanograins. These energies exceed by a
considerable amount those of class I and II. Nonetheless, multigrain
nanograins are non-equilibrium structures that are likely to exist under
experimental conditions. In addition, the recent simulations of the structure
of ultra-small Al clusters (Manninen {\em et al.}, in Ref.
\onlinecite{note-021}) show that complex, multigrain structures are
competitive with those of class I and II.  Spherical grains, on the other
hand, correspond to an idealization of realistic metal grains and are
presented here only for comparison purposes.

%
\begin{table}[tb]
\caption{\label{Table_1} Total energy of multigrain and spherical Ag grains vs
  number of atoms $N$. Here, $\varepsilon_B=2.85\,{\rm eV}$ is the bulk cohesive energy.}
\begin{tabular}{cccccc}
\hline\hline
 & \multicolumn{4}{c}{$({\cal E}-\varepsilon_B N)/N^{2/3}$ ({\rm eV})} \\
\cline{2-5}
$N$ &  225 & 459 & 783 & 1289  \\
\cline{2-5}
Multigrain & 1.93 & 1.94 & 1.98 &  & \\
{\it Spherical} & 2.06 & 2.06 & 1.99 & 2.04 & \\
\hline\hline
\end{tabular}
\end{table}

%
%
\section{Vibrational properties}

We show that nanograin morphology strongly affects the vibrational properties.
By using the methods described in the Appendix, we analyze the effect of
nanograin's morphology on the phonon frequency spectrum, vibrational density
of states (DOS), and phonon localization. First, we describe general features
of phonon spectra followed by a comparison between class I and
spherical grains with class II and multigrains. Second, we contrast the DOS of
{\sf TO} with class II and III nanograins.  Finally, we focus on N=459
nanograins in class I and III to illustrate important features of the phonon
localization.

\subsection{Phonon frequencies}
\label{Phonon_Freq}

{\em General features.} The finite size of the grains lead to discrete phonon
spectra. The acoustic gap, which is defined as the value of the lowest phonon
frequency, decreases with size due to confinement.  Our simulations show that
the acoustic gap roughly scales as $N^{-1/3}$, as predicted by elasticity
theory (ET) for a finite grain,\cite{landau_book} and its magnitude depends on morphology.
Each grain in class I as well as icosahedra ({\sf Ih}) have a nearly-radial,
breathing mode. In contrast, this mode does not appear in Mark's decahedra
({\sf M-Dh}) or multigrains (class III). When compared with predictions of
elasticity theory for a solid sphere with the same volume as the
nanograins, our simulations lead to frequencies of the breathing mode ($R_1$)
and acoustic gap about $10\%$ smaller. For $R_1$, ET predicts
$\nu^{N}_{\rho}=2.825\,(c_l/2\pi R)$, where $R$ is the radius of the sphere
and $c_l=3686\,{\rm m/s}$ is the average longitudinal sound velocity of bulk
silver.\cite{anderson_chapter} The ET-acoustic-gap equals
$0.401\,\nu^{N}_{\rho}$ with 5-fold
degeneracy.\cite{patton_PRB_2003,tamura_JPC_1982}

%
\begin{figure}[tb]
\includegraphics[width=8.5cm]{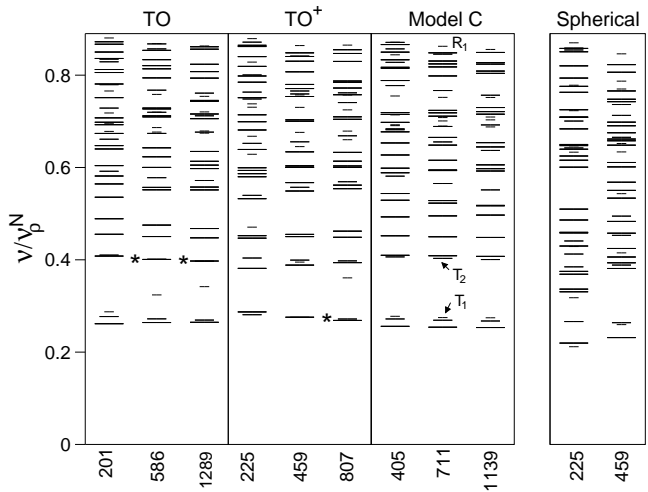}
\caption{\label{Fig_3} Single-crystalline Ag grains' scaled phonon frequencies $\nu$
  up to the first nearly-radial, breathing mode ($R_1$).
  $\nu^{N}_{\rho}=2.825\,(c_l/2\pi R)$ (see Sec.  \ref{Phonon_Freq}).
  Frequencies are either nondegenerate (short dash) or 2-fold (medium dash)
  and 3-fold (long dash) degenerate frequencies are present. In {\sf TO},
  ${\sf TO^{+}}$ and {\sf Model C}, the low-lying phonon frequency structure
  is qualitatively similar except for $T_1$; a nearly-torsional, edge
  vibration (see Fig.  \ref{Fig_4}). $T_2$ labels a 2-fold degenerate twist
  mode. Asterisks (*) indicate nearly degenerate doublet-triplet clusters.
  Surface protrusions yield different low-frequency phonon structure in
  spherical grains.}
\end{figure}
%

%
{\em Class I and spherical nanograins.}  For {\sf TO}, ${\sf TO^+}$, and {\sf
  Model C} in class I, and for the spherical grains in class III, Figure
\ref{Fig_3} shows phonon frequencies up to the first nearly-radial, breathing
mode. Main features are the following. (i) All grains show frequencies that
are either non-degenerate or 2-fold and 3-fold degenerate.
(ii) The five lowest frequencies split into a triplet and a doublet; a gap
separates this cluster from other frequencies. This triplet-doublet cluster
originates from symmetry breaking of the ET-acoustic-gap quintet. The size
dependence of the small split-off separation between the triplet and doublet
depends on the morphology. It decreases for {\sf TO}, and in ${\sf TO^+}$,
N=459 the triplet and doublet even switch order. {\sf Model C} shows a more
remarkable size dependence: The split-off separation remains nearly constant
and similar to that of N=201 {\sf TO} grain. We attribute this behavior to the
fixed number of atoms (3) in the [111]-facet boundaries.
(iii) In ${\sf TO^+}$ and in 586 and 1289 {\sf TO} grains, a distinguishable,
second group of frequencies appears; a triplet precedes a doublet and two
triplets follow.  The separation between multiplets in these complexes also
depend upon the size of the nanograin. In contrast, no visible gap separates
these frequencies in {\sf Model C} and 201, {\sf TO}.  At frequencies higher
than $0.5\nu^{N}_{\rho}$ no sizeable gaps appear in any grain.
(iv) In all grains, near to the lowest triplet-doublet, we find a
low-frequency nearly-torsional edge mode $T_1$; morphology dramatically
affects the size dependence of its frequency (Fig. \ref{Fig_3}).  In {\sf TO},
$T_1$'s frequency remains almost constant while in ${\sf TO^+}$ it decreases
with the surface. In {\sf Model C}, it scales as $\nu^{N}_{\rho}$ and is {\em
  locked} with the lowest five frequencies.

The ideal spherical nanograins present phonon frequencies that are
nondegenerate as well 2- and 3-fold degenerate. Contrary to class I grains,
the low-lying frequencies present a less clear pattern. These grains, N=225
and 459, may be viewed as the {\em decorated} version of the 201 {\sf TO}
(c.f. Fig. \ref{Fig_1}) and 405 ${\sf TO^+}$, respectively.  The coupling
between the vibration of the protrusions and the atoms in the faceted grain
that they decorate reduces the acoustic gap. This coupling also destroys the
similarities between the frequency structure in 201 and 405. (See Fig.
\ref{Fig_2}.)

%
\begin{figure}
\includegraphics[width=8.5cm]{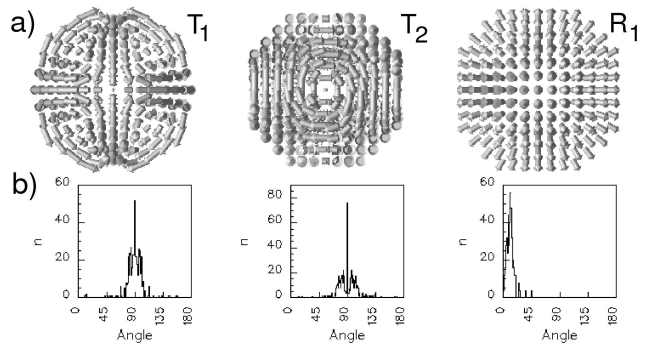}
\caption{\label{Fig_4}(a) Displacement vectors of selected phonon 
  modes in the N=586, {\sf TO} nanograin: (i) $T_1$, a non-degenerate mode in
  which the atoms surrounding the edges have a significant amplitude, (ii)
  $T_2$, one of the doubly-degenerate twist modes, and (iii) $R_1$, the first
  nearly-radial, breathing mode. (b) Distribution functions ($1^{\rm o}$ bin
  size) of the angle between the equilibrium vector position of an
  atom---measured from the center of the grain---and its displacement vector.
  The distributions peak around 90$^{\rm o}$ for $T_1$ and $T_2$, and
  10$\,^{\rm o}$ (instead zero) for $R_1$, respectively. For larger grains the
  distributions peak around 2-4$\,^{\rm o}$.}
\end{figure}
%

%
For N=586 {\sf TO} nanograin, Fig. \ref{Fig_4}(a) shows representations of the
displacement vectors of 3 selected phonon modes. As mentioned above, $T_1$ is
nearly torsional and the atoms surrounding the edges have the higher
amplitude. $T_2$ is one of the two nearly-torsional, 2-fold generate twist
mode. In contrast to $T_1$, $T_2$'s frequency depends less strongly on
morphology (see Fig. \ref{Fig_3}). $T_1$ and $T_2$ also appear in spherical
grains.  $R_1$ is the non-degenerate, nearly-radial breathing mode.  It should
be noted that {\sf TO} and {\sf Ih} grains have another nearly-torsional mode
$T_3$ (not shown), which is an excited, non-degenerate twist mode.  This mode
does not appear in any other single-crystalline grain or Mark decahedron, as
only {\sf TO}'s and {\sf Ih}'s have the same number of atoms at the every
[111]-facet boundary. [c.f. Figure \ref{Fig_1}(a) and (b)]

For $T_1$, $T_2$, $R_1$, figure \ref{Fig_4}(b) shows histograms of the angles
between the displacement vector of atom i ($\vec{u}^i$) and its equilibrium
position ($\vec{R}^i$); measured from the center of the grain. [The angle is
defined mod($\pi$).]  The purpose is to show the extent to which the
crystalline field and edges of the grains couple angular and radial degree's
of freedom. Note that in an ideal elastic sphere, torsional and breathing
modes have angular- and radial-only displacements,
respectively.\cite{patton_PRB_2003}
The values for the average angle $\overline{\theta}$ and the standard
deviation $\sigma$ of the histograms in Fig. \ref{Fig_4} are the following:
$\overline{\theta}_{T_1}=89.51^{\rm o}$, $\sigma_{T_1}=12.53^{\rm o}$;
$\overline{\theta}_{T_2}=89.95^{\rm o}$, $\sigma_{T_2}=14.46^{\rm o}$; and
$\overline{\theta}_{R_1}=10.64^{\rm o}$, $\sigma_{R_1}=5.96^{\rm o}$.  In
general, as size increases, for nearly torsional and radial modes,
$\overline{\theta}$ approaches $90^{\rm o}$ and $0^{\rm o}$, respectively; and
$\sigma$ decreases.

%
\begin{figure}[htb]
\includegraphics[width=8.5cm]{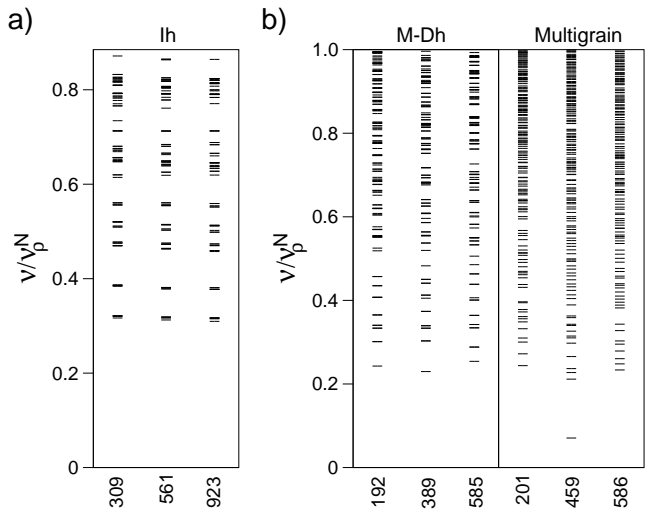}
\caption{\label{Fig_5}(a) Equivalent to Fig. \ref{Fig_3} but for twined
  icosahedral grains. Strain leads to clusters of non-degenerate frequencies,
  and acoustic gaps that are bigger than in single-crystalline grains and
  Mark's decahedra. (b) Mark's decahedra and multigrain's phonon frequencies
  up to $\nu^{N}_{\rho}$. These two nanograins {\em do not have} nearly-radial
  phonon modes. Like {\sf Ih}'s, {\sf M-Dh}'s show non-degenerate frequencies
  in this range. In multigrain, grain boundaries and surface disorder lead to
  non-degenerate frequencies and significantly smaller acoustic gap.}
\end{figure}
%

{\em Class II and multigrain nanograins.}  Figure \ref{Fig_5}(a) shows the
phonon frequencies up to the first nearly-radial, breathing mode of twined
icosahedra ({\sf Ih}). Panel (b) shows phonon frequencies up to
$\nu^{N}_{\rho}$ for Mark's decahedra ({\sf M-Dh}) and multigrain samples.
(See Fig. \ref{Fig_1}.) Prominent features are the following:

(i) Strain in the icosahedra leads to clusters of non-degenerate frequencies.
At higher frequencies a few 2-fold degeneracies appear.  The acoustic gap is
higher than in class I and spherical grains.  It also has a triplet-doublet
structure with higher split-off and no degeneracies.
(ii) Mark's decahedra, like {\sf Ih}'s, only show a few doubly degenerate
frequencies over the entire spectrum. The acoustic gap is similar in magnitude
to the gaps of class I and spherical grains. However, there is no resemblance
of the triplet-doublet structure. More importantly, Mark's decahedra do not
have breathing modes. The small [111]-oriented dents (see Fig. \ref{Fig_1})
may be responsible for these features.
(iii) Multigrains only have non-degenerate frequencies.  Like {\sf M-Dh},
multigrain do not have breathing modes. Surface disorder and defects reduce
the acoustic gap. (Compare Fig. \ref{Fig_3} and \ref{Fig_4}.) The lowest-lying
modes are localized on the surface of the grain; however, the degree of
localization is sample (disorder) dependent. In addition, the more localized
the mode the softer its frequency.  The visible gap between the lowest allowed
phonon mode and the first {\it excited} vibration in the N=459 nanograin is
disorder sensitive.

\subsection{Vibrational density of states}

\begin{figure}[htb]
\includegraphics[width=8.5cm]{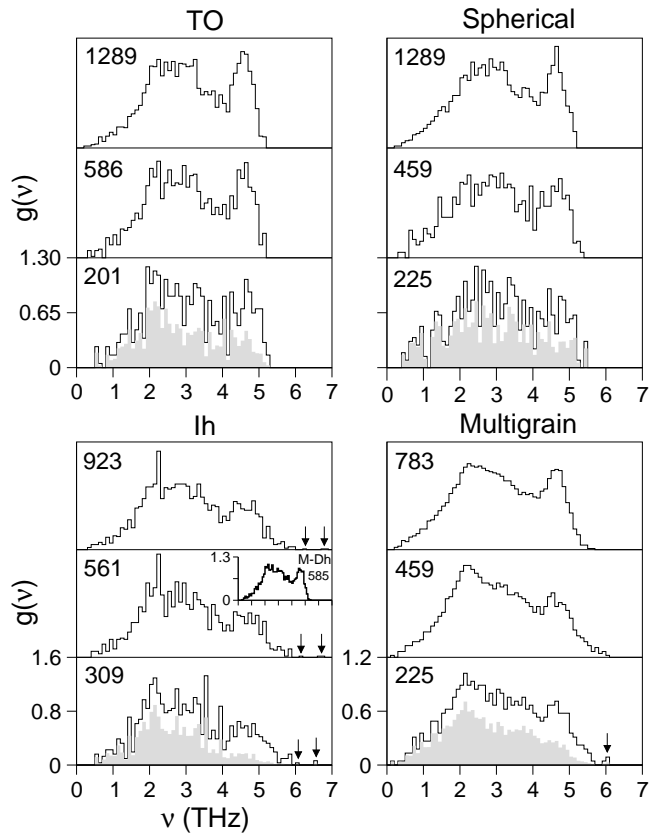}
\caption{\label{Fig_6} Morphology and size effects on vibrational density 
  of states $g(\nu)$. Shaded histograms show the partial density of states due
 to surface atoms. Bin size is $0.1\,{\rm THz}$. At low and high frequencies,
  $g(\nu)$ depends significantly on geometrical details of the nanograins.
  Density of states at $\nu<2\,{\rm THz}$ is higher in spherical grains than
  in {\sf TO}.  For faceted and {\sf Ih} grains, the longitudinal-phonon (LP)
  peaks around $4.7\,{\rm THz}$ have nearly the same amplitude regardless of
  size. In contrast, LP's amplitude grows with grain size in multigrain and
  spherical grains. Grain boundaries in multigrain and twined {\sf Ih} lead to
  a frequency tail beyond the maximum value in single-crystalline grains. No
  tail is seen in Mark's decahedra, whereas the LP peak is visible. [See inset
  in panel {\sf Ih}, 561 for $g(\nu)$ of a 585 {\sf M-Dh}.] Arrows indicate
  features arising from phonons strongly localized in the center of the
  grains.}
\end{figure}

Figure \ref{Fig_6} shows the effects of morphology and size on the vibrational
density of states (DOS) per atom $g(\nu)$. In these results, each $g(\nu)$
histogram has a 0.1$\,{\rm THz}$ bin.\cite{note-011}

%
%

%
\noindent At low and high frequencies, $g(\nu)$ depends strongly on {\em {\sf TO}
  grains} size. DOS in {\sf TO} grains increase up to $2\,{\rm THz}$, which
correspond to a maximum in bulk DOS, originated by transversal phonons (TP).
The longitudinal-phonon peaks around $4.7\,{\rm THz}$ have nearly the same
amplitude for different sizes. Other class I grains (c.f. Sec.
\ref{Section_2}) show a similar behavior. DOS extends to slightly higher
frequencies for smaller grains.
{\em Icosahedra and Mark's decahedra.---}In icosahedra, strain and twinning
remarkably smear the LP peak in the DOS. Its intensity is lower than the TP
peak about $2\,{\rm THz}$, and remains roughly unchanged with increasing size.
In addition, there is a frequency tail beyond the highest frequency in {\sf
  TO} and spherical beyond $6\,{\rm THz}$, the DOS show small features (arrows
in Fig.  \ref{Fig_6}, {\sf Ih}) that originate in a four phonon modes strongly
localized at the center of the grains. These phonon modes consist of
displacement of a few (5-40) atoms along the grain-boundary interfaces that
meet at the center of grain. Removal of the center atom of the icosahedra
results in the disappearance of these features in the density of states.
Mark's decahedra show a DOS significantly different from icosahedra; see the
inset in Fig. \ref{Fig_6}, panel {\sf Ih}, 561. Although {\sf M-Dh}'s are
twined and strained, an LP peak is visible and its intensity remains nearly
unchanged with size (not shown).  More important, Mark's decahedra do not show
a high-frequency tail in DOS.
{\em Multigrain and {\it spherical} nanograins.---}In spherical grains,
protrusions increase $g(\nu)$ below 2 and above 5$\,{\rm THz}$ relative to
faceted grains. Although less prominently that in multigrains, the amplitude
of LP peaks also increases and narrows with size. In multigrain, (i) the
amplitude of the LP peaks grows and narrows with size. Remarkably, Pasquini
{\em et al}.\cite{pasquini_PRB_2002} observed a similar build-up of longitudinal
phonons of nanocrystalline $\alpha$-Fe.\cite{note-11} (ii) Similarly to {\sf
  Ih}, a frequency tail appears at high frequencies. (Arrows also point to
features arising from inner atoms.)

Finally, let us mention that Derlet and Swygenhoven\cite{derlet_PRL_2004} have
recently shown that the features at higher frequencies in the DOS of
nanocomposites originate from vibrations strongly localized in grain
boundaries. The frequency tail that appears in $g(\nu)$ for {\sf Ih} and
multigrain is an intrinsic feature of nanoscopic metals with grain boundaries.
This remarkable finding reveals the strong connection between extended
nanocrystalline materials and finite size nanoscopic grains.

%
%
\subsection{Phonon localization}
\label{Local}

The participation ratio $N_{eff}(\nu)$ (see Appendix) indicates the effective
number of atoms participating in a phonon with frequency $\nu$. Thereby,
$N_{eff}(\nu)$ measures the degree of phonon localization. In a sample with N
atoms, $N_{eff}/N\simeq 1/2$ for extended, wave-like phonon
modes.\cite{note-001} Note that when we identify a phonon mode as being
localized [$N_{eff}/N\ll 1/2$] we do not imply necessarily that the
amplitude of the atomic vibrations decays away exponentially from a particular
atomic site, as it is the case for Anderson-localized vibrational
states.\cite{maynard_RMP_2001} In addition, while Anderson localization takes
place in disordered systems, the term localization in the context of our
results is broader as grains have a surface and, therefore, phonon
localization---i.e., phonon modes with a small number of atoms involved in the
vibration---takes place even in crystalline grains.

%
\begin{figure}
\includegraphics[width=8.5cm]{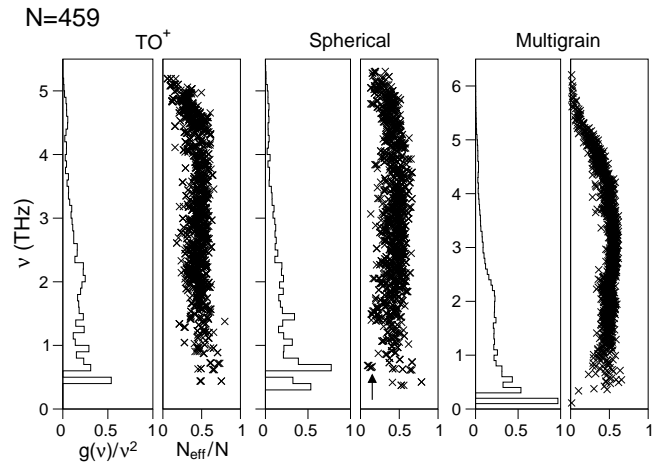}
\caption{\label{Fig_7} Morphology effects on the reduced density of states 
  $g(\nu)/\nu^2$ and the effective number of atoms $N_{eff}(\nu)$ in a phonon
  of frequency $\nu$. For wave-like modes, $N_{eff}/N=1/2$. Only multigrain shows a
  $\nu^{2}$ distribution from 1 to slightly over 2$\,{\rm THz}$. In all
  grains, edges and vertices significantly enhance $g(\nu)/\nu^2$ below
  1$\,{\rm THz}$.  An arrow indicates modes heavily localized on the
  protrusions of the spherical grain, which coincide with the peak around
  $0.5\,{\rm THz}$. In the multigrain, grain boundaries and irregular surfaces
  are responsible for the reduced fluctuations shown in $N_{eff}/N$.}
\end{figure}

Figure \ref{Fig_7} shows $N_{eff}(\nu)$ scaled by N, and the reduced
vibrational density of states $g(\nu)/\nu^2$ for N=459 ${\sf TO^+}$,
spherical and multigrain. The central results can be summarized as follows.
(a) General features of $N_{eff}$ are representative for all nanograins: (i)
High-frequency phonons become more localized in both single-crystalline and
multigrain specimens. In the latter, grain boundaries lead to a higher degree
of localization. This is similar to what occurs in nanocrystalline
materials;\cite{derlet_PRL_2004} a result that shows that nanograins capture
some of the physics of their massive, nanocrystalline
counterparts.\cite{note003} (ii) At low frequencies, the more localized the
vibrations the higher the enhancement of DOS. In particular, spherical grains
show a peak (arrow, Fig. \ref{Fig_7}) around $0.6\,{\rm THz}$ due to localized
phonon modes at the surface protrusions. In turn, multigrain shows a similar
enhancement due to phonons localized around vertex vacancies and in atoms in
the vicinity of grain boundaries. In particular, two low-frequency modes at
$0.1\,{\rm THz}$ and $0.3\,{\rm THz}$ that involve $\sim 5$ and $50$ atoms,
respectively, are responsible for the low-frequency peak in $g(\nu)/\nu^2$.
Faceted single-crystal and twined samples (not shown) show a lesser
enhancement, arising only from the edges and vertices. (iii) $N_{eff}$
significantly fluctuates around $N/2$ for single-crystal samples, twined {\sf
  Ih} and {\sf M-Dh}; whereas in multigrain, irregular shape, surface
disorder, and defects reduce these fluctuations.

(b) Two prominent features appear in $g(\nu)/\nu^2$: (i) The nearly quadratic
behavior of $g(\nu)$ from $1-2.2{\rm THz}$ for multigrain. The frequency
extent of this Debye-like [$g(\nu)\sim\nu^2$] region depends on the nature of
disorder and defects---multiple grain boundaries, stacking faults, and vertex
vacancies. Such Debye dependence has been observed in iron nanocomposites by
Pasquini and coworkers.  \cite{pasquini_PRB_2002} The Debye constant derived
from the quadratic interval is nearly twice as big as the bulk value for Ag.
Remarkably, such enhancement of the Debye constant has also been observed by
Fultz and coworkers in nanocrystalline iron.\cite{fultz_PRL_1997} It should be
noted that while we have found Debye-like behavior in a few nanograins of
different sizes, the statistics we have is limited due to the intrinsic
difficulty in generating competitive (Sec. \ref{Section_2}) multigrain samples
using long molecular-dynamics simulations. (ii) The dramatic correlation
between $N_{eff}(\nu)$ and $g(\nu)/\nu^2$, which shows that the degree of
phonon localization in nanoscopic metal grains can be derived from a
measurement of their vibrational density of states, after dividing by $\nu^2$.

%
\begin{figure}
\includegraphics[width=8.5cm]{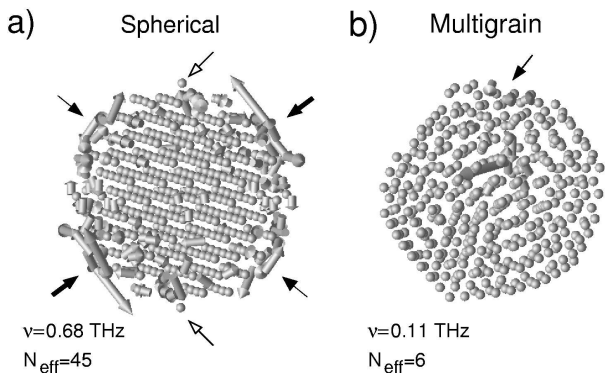}
\caption{\label{Fig_8} Displacement vectors for two selected low-frequency,
  localized phonon modes in N=459 spherical (a) and multigrain (b) samples.
  For each mode, we also show the frequency ($\nu$) and $N_{eff}$ (Sec. \ref{Local}
  and App. \ref{Methods}). In (a), the atoms with significant displacement-vector
  amplitude are those located on the 6-atom protrusions (indicated with solid
  arrows) that decorate the [111] facets of the underlying {\sf TO} grain
  [Fig. \ref{Fig_1}(c)]. Single-atom protrusions that decorate the [001]
  facets have negligible displacement-vector amplitude (open arrow). In (b),
  the phonon mode involves a very low number of atoms ($N_{eff}=6$) that are
  around a vertex vacancy (indicated with an arrow) and grain boundaries.}
\end{figure}

Finally, Figs. \ref{Fig_8}(a) and (b) show, respectively, displacement vectors
for two low-frequency, localized ($N_{eff}/N\ll 1/2$) phonon modes in N=459
spherical and multigrain samples. These phonon modes are among those
responsible for the low-frequency peaks in $g(\nu)/\nu^2$ (Fig. \ref{Fig_7}).
We find the following. (i) In the spherical grain, the phonon mode is one of a
three-fold degenerate set with frequency $\nu=0.68\;{\rm THz}$ and
$N_{eff}/N=0.098$.  This mode is mostly localized in the 6-atom protrusions
that decorate the hexagonal, [111] facets of the underlying {\sf TO} grain
(Fig. \ref{Fig_1} and Sec.  \ref{Section_2}). The amplitude of the
displacement vectors of the atoms in the protrusions is {\em different}
depending on which facet the protrusion is located, as indicated by solid,
thick- and thin-stem arrows in Fig.  \ref{Fig_8}(a), whereas displacement
vectors of atoms in diametrically opposed protrusions have the same amplitude
but different phase. Note that in this phonon mode [Fig.  \ref{Fig_8}(a)] the
single-atom protrusions located on top of [001] facets have negligible
displacement-vector amplitude (open arrows).
(ii) The phonon mode in the multigrain sample [Fig. \ref{Fig_8}(b)] is the
lowest allowed phonon.  Its frequency is $\nu=0.11\;{\rm THz}$ and
$N_{eff}/N=0.013$.  The displacement vectors with significant amplitude locate
around a vertex vacancy, indicated with an arrow, and along grain boundaries.

\section{Summary}

To understand in detail the vibrational properties of nanoscopic metal grains
we performed atomistic simulations of the phonon spectra---frequencies and
displacement vectors---of silver grains with realistic morphologies.  Our
results show that morphology introduces a high degree of complexity into the
phonon spectra, total and partial vibrational density of states, and phonon
localization. The most prominent features are the following.
(a) Low-energy, single-crystalline grains present nearly-pure torsional and
radial phonon modes. The grains' sharp edges and atomic lattice split off the
acoustic gap quintet predicted by ET into a doublet and triplet, with a
magnitude that depends on the relative number of atoms at the boundary of
different facets. When compared to faceted grains of the same size,
high-energy, spherical models that present regular protrusions on the surface
have a smaller acoustic gap and a higher total DOS at frequencies $\nu<2\,{\rm
  THz}$.
(b) Twined icosahedra also have a breathing mode.  Strain in these grains
leads to a higher acoustic gap, and a high-frequency tail in the DOS that
originates from core atoms. This tail extends beyond the highest frequency in
single-crystalline grains. Remarkably, {\em neither} twined and strained
Mark's decahedra have a breathing mode nor do they exhibit a high-frequency
tail on the DOS.
(c) Nanograins with grain boundaries and surface disorder do not have
degenerate frequencies and the acoustic gap is significantly reduced. These
nanograins are the only ones that exhibit low-frequency $\nu^2$ DOS in the
interval 1-2${\rm THz}$.  The extent of this region depends on the nature of
disorder.

These predictions, while illustrated here for silver, should be valid for
other nanoscopic metal grains. Our simulations show the complexity of the
vibrational properties of metal nanograins, and point out similarities
between the vibrational density of states of metal nanograins with grain
boundaries and surface disorder and that of massive nanocrystalline samples.
Finally, this work is a necessary step toward understanding the size-dependent
electron-phonon coupling in nanoscopic metals.

\section*{Acknowledgments}

Funds from U. S. DOE grant DE-F G02-99ER45795 and computational resources from
the National Center of Supercomputer Applications (NCSA) made this research
possible.  National Science Foundation grant NSF-DMR-03-25939-ITR supports the
Materials Computation Center. The authors thank Professor Jian-Min Zuo (UIUC)
for fruitful discussions.  This research started at the Materials Computation
Center (MCC).  One of the authors (G.A.N.)  acknowledges its hospitality, and
the opportunity to participate in the MCC Visitor Program.

\appendix
%
%
\section{Methods}
\label{Methods}

{\em Nanograin's structure and energetics.}  The literature contains extensive
studies regarding the energetics of metal clusters.  In this work, we consider
known geometries and perform total energy relaxations in Ag nanograins. We
also perform extensive simulated annealing to get complex, partially
disordered morphologies.\cite{noteonMD} To calculate the total energy $\cal
{E}$ we employ Daw and Baskes' embeded atom method\cite{daw_PRB_1984} (EAM).
We use Voter and Chen's parametrization of Ag potential.

{\em Vibrational properties.} 
%
Once we generate stable nanograin morphologies we calculate phonon frequencies
$\nu$ and displacement vectors $\{\vec{u}^{\,i}_{\nu}\}$---where $i$ labels an
atom located at equilibrium position $\vec{R}^{\,i}$ in the grain---by direct
diagonalization of the dynamical matrix $D^{ij}_{\alpha\beta}=\partial^2{\cal
  {E}}/\partial{R^{i}_{\alpha}}\partial{R^{j}_{\beta}}$.
$\vec{R}^{i}_{\alpha}$ is a cartesian component ($\alpha,\beta=x,y,z$) of
$\vec{R}^i$.  Note that we consider free-standing nanograins, hence, the first
six phonons correspond to rigid-body translations and rotations along and
about the principal axis, respectively.  These frequencies do not enter in our
analysis.

%
The vibrational density of states (DOS) $g(\nu)=\sum^{N}_{j=1}\,g_j(\nu)$,
where $g_j(\nu)\Delta\nu=(1/N)\sum_{\nu}n(\nu)\sqrt{M}|\vec{u}^{\,j}_{\nu}|$
is the partial (local) vibrational density of states at atomic site j;
$\vec{u}^{\,j}_{\nu}$ is the displacement vector of atom $j$; $n(\nu)$ is the
number of phonons between $\nu$ and $\nu+\Delta\nu$; and $M$ is the mass of a
Ag atom.

%
The equality $N_{eff}(\nu)=M^2_1/M_2$ defines the participation ratio.
\cite{jin_PRB_1993,nagel_PRL_1984,dean_RMP_1972} Here,
$M_p=\sum^{N}_{i}[\varepsilon_i(\nu)]^p$ are momenta of the {\it mean} kinetic
energy $\varepsilon_i(\nu)$ of atom $i$. By using the calculated displacement
vectors of atom $i$ in a phonon with frequency $\nu$ one readily calculates
$\varepsilon_i(\nu)$.


\begin{thebibliography}{100}
\bibitem{note-0000} One may argue that such grains would be energetically
  unfavorable; however, these are more competitive than the simple models;
  see Sec. \ref{Section_2}.
\bibitem{noteonMD} Our simulated annealing is based on molecular dynamics.  We
  use Verlet's algorithm with time steps from 1-2$\,{\rm fs}$, and run
  multi-million-step simulations in the microcanonical ensemble. Typically, we
  cool down a molten droplet down to 300-500$\,{\rm K}$ at rates from
  20-50$\,{\rm K/ns}$.  Then, we relax the atomic positions to minimize ${\cal
    E}$.  Although we use high cooling rates, they are practical and fall
  within the range normally used in the literature.
\end{thebibliography}
\end{document}